\documentclass[dvips]{article}
\usepackage{icrctc07,multirow}
\title{STACEE Observations of 1ES 1218+304}
\shorttitle{STACEE Observations of 1ES 1218+304}
\authors{R. Mukherjee,$^a$ N. Akhter,$^a$ J. Ball,$^b$ J. E. Carson,$^{b,1}$ C. E. Covault,$^c$ D. D. Driscoll,$^c$ P. Fortin,$^a$  
D. M. Gingrich,$^{d,e}$ D. S. Hanna,$^f$
A. Jarvis,$^b$ J. Kildea,$^{f,2}$ T. Lindner,$^{f,3}$ 
C. Mueller,$^f$ 
R. A. Ong,$^b$ 
K. Ragan,$^f$ D. A. Williams,$^g$ J. Zweerink$^b$}
\shortauthors{R. Mukherjee et a.}
\afiliations{        (a) Dept. of Physics \& Astronomy, Barnard College, Columbia University, New York, NY 10027\\
        (b) Department of Physics and Astronomy, University of California at Los Angeles, Los Angeles, CA 90095\\
        (c) Department of Physics, Case Western Reserve University, 10900 Euclid Ave., Cleveland, OH, 44106\\
        (d) Centre for Particle Physics, University of Alberta, Edmonton, AB T6G 2G7, Canada \\
        (e) TRIUMF, Vancouver, BC V6T 2A3, Canada\\
        (f) Department of Physics, McGill University, 3600 University Street, Montreal, QC H3A 2T8, Canada\\
        (h) {Santa Cruz Institute for Particle Physics, Univ. of California at Santa Cruz, 1156 High Street, Santa Cruz, CA 95064}\\ 
        (1) Current Address: Stanford Linear Accelerator Center, MS 29, Menlo Park, CA 94025\\
        (2) Current Address: Fred Lawrence Whipple Observatory, Harvard-Smithsonian Center for Astrophysics, Amado, AZ 85645\\
        (3) Current Address: Department of Physics and Astronomy, University of British Columbia, Vancouver, BC V6T 1Z1, Canada\\
}
\email{muk@astro.columbia.edu}

\abstract{We present the analysis and results of recent high-energy gamma-ray
observations of the high energy-peaked BL Lac (HBL) object 1ES 1218+304 with
the Solar Tower  Atmospheric Cherenkov Effect Experiment (STACEE). 1ES
1218+304 is an X-ray bright HBL at a redshift
z=0.182. It has been predicted to be a gamma-ray emitter above
100 GeV, detectable by ground-based Cherenkov telescopes. Recently,
this source has been detected by MAGIC and VERITAS, confirming these
predictions. STACEE's sensitivity to astrophysical sources at energies
above 100 GeV allows it to explore high energy sources such as X-ray
bright active galaxies and gamma-ray bursts. We present results
from STACEE observations of 1ES 1218+304 in the 2006 and 2007 observing
seasons.
}

\begin{document}
\maketitle

\section{Introduction}
Active galaxies of the ``blazar'' class include BL Lac objects and 
flat-spectrum radio quasars (FSRQs), and are characterized by 
non-thermal continuum emission that extends from radio to high energy gamma rays. The spectral energy distributions (SEDs) of these sources 
typically have two broad peaks, one at low energies (radio
to X-ray) and the other at higher energies (keV to TeV). In the
framework of relativistic jet models, 
these objects are highly beamed sources, emitting plasma in
relativistic motion (e.g. \cite{Urry&Padovani1995}). In blazar SEDs, 
the low energy peak is explained as synchrotron emission from high
energy electrons in the jet, while the high energy peak is probably due
to inverse Compton emission. Several competing ``leptonic'' and
``hadronic'' jet model explanations exist for the high energy emission (e.g. see
\cite{Boettcher2002} \& \cite{Muckeetal2003} for reviews), and further broadband
observation of blazars are needed to distinguish between these models. 

Since the
discovery of blazars as high energy gamma-ray sources by the Energetic
Gamma-Ray Experiment Telescope (EGRET) on board the {\sl Compton Gamma
  Ray Observatory} (CGRO) \cite{Hartman1999} and the first detection
of a TeV ($10^{12}$ eV) 
blazar by the Whipple Observatory (Mrk 421 \cite{Punch1992}), 
the search has been on for more TeV blazars. The number count of TeV
blazars is growing, with the advent of new 
generation atmospheric Cherenkov
telescopes (ACTs) \cite{Hessblazar19xx}. To date, almost all confirmed blazars
detected at TeV energies are high-frequency-peaked BL Lac objects
(HBLs), as opposed to quasars that constitute the majority of the
EGRET detections. HBLs are a sub-class of BL Lac objects, 
characterized by lower luminosity than FSRQs and a synchrotron peak
in the X-ray band \cite{Urry&Padovani1995}. 
Blazars are categorized into different sub-classes based on the peak
frequencies and the relative power in the low and high energy peaks of
their SEDs \cite{Fossati1998}. Given the high synchrotron peak
frequencies of HBLs,
indicating the presence of high energy electrons, these
sources have been predicted to be good candidates for TeV emission,
based on synchrotron self-Compton (SSC) emission models
\cite{Costamante&Ghisellini2002} as well as hadronic models
\cite{Mannheim1993}. Several of the ``extreme'' synchrotron BL Lacs
\cite{Costamante2001} have been detected at TeV energies, confirming
these predictions. 

1ES 1218+304 is an X-ray bright (flux at 1 keV $> 2$ $\mu$Jy) HBL,
categorized as an ``extreme'' BL Lac, and predicted
to be a TeV source. The source was recently detected by both MAGIC
\cite{Albert2006} and VERITAS \cite{Fortin2007}, at
energies $>100$ GeV. At a redshift of $z=0.182$, 1ES 1218+304 is one
of the most distant blazars detected to date. The source was never
detected by EGRET, indicating that ACTs are sensitive to
a different population of gamma-ray blazars than EGRET. 

The Solar Tower Atmospheric Cherenkov Effect Experiment (STACEE) is a
ground-based experiment that is sensitive to gamma rays above 100
GeV. STACEE observations of AGN are motivated by the need to understand particle
acceleration and emission mechanisms in blazars, as well as their
interaction with the extragalactic background light (EBL). Despite what is already known, a great deal
remains to be discovered regarding the physics of blazars.  STACEE's
extragalactic observing program has included both HBLs, as well as
LBLs \cite{Mukherjee2006, Mukherjee2005}. Recent observations of
1ES 1218+304 with STACEE were motivated by the detection of TeV
emission from the source by MAGIC, providing further evidence that
X-ray bright HBLs tend to be strong VHE sources. STACEE observations
of 1ES 1218+304 were carried out in the 2006 and 2007 observing
seasons. In this paper we present a summary of STACEE
observations of the HBL 1ES 1218+304. 

\section{The STACEE Detector}

STACEE is a shower-front sampling Cherenkov detector that operates with 
an energy threshold of about 100 GeV. It uses 
64 large, steerable mirrors (heliostats) at the National  
Solar Thermal Test Facility (NSTTF) near Albuquerque, NM, USA to
collect Cherenkov light from extensive air showers. 
The large light-collection
area ($\sim 2400$ $\rm m^2$) gives it a lower energy threshold than
all but the newest imaging Cherenkov detectors. STACEE uses secondary
mirrors, in the central receiver tower to focus Cherenkov light
reflected by the heliostats onto photomultiplier tubes (PMTs), with a
one-to-one mapping between the heliostats and the PMTs (Figure 1 shows
a schematic diagram). The 64 channels are grouped into eight clusters
of eight channels each for triggering, and the two level trigger
requires that all events trigger at least five tubes in each of at
least 5 clusters. 
STACEE has been
operating as a complete detector  since 2001 and has observed numerous
astrophysical sources \cite{Oseretal2001, Carsonetal2007}. Details of
the STACEE detector and operations are given elsewhere
\cite{Gingrichetal2004}. 

\begin{figure}[b!]
\begin{center}
\includegraphics*[width=0.48\textwidth]{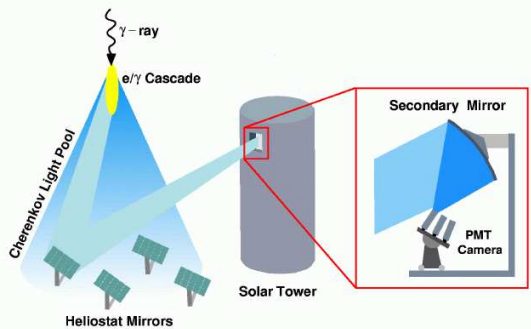}
\end{center}
\caption{The shower-front sampling technique as employed by
  STACEE. Cherenkov light from air shower cascades is reflected by the
  heliostat mirrors onto secondary mirrors in the receiver tower,
  which in turn concentrate the light onto a bank of photomultiplier
  tubes. }\label{fig1}
\end{figure}

\section{Observations \& Data Analysis} 

STACEE observed 1ES 1218+304 from February to June, 2006 and February
to May, 2007. 
Data taken with STACEE consist of 28-minute paired ON-source and OFF-source
observations, the latter taken to determine the background. Table 1
summarizes the STACEE data and the total livetime on 1ES 1218+304. 

The STACEE data set was analyzed off line to remove data taken in unfavorable
weather conditions or with detector malfunctions (e.g. malfunctioning 
heliostats, high voltage trips, etc.), eliminate biases in the
trigger rates and increase the sensitivity of the instrument. STACEE
data cleaning criteria or {\sl cuts} are described elsewhere
\cite{Brameletal2005}. In 
addition, cosmic ray background suppression techniques were 
applied to the data, as described in \cite{Kildea2005a,
  Kildea2005b}. 

During the 2006 and 2007 observing seasons, a total of
152 ON-OFF pairs (70.9 hr) were taken on 1ES 1218+304. After several
standard data quality cuts, the total 
number of hours on source was reduced to 28.3 hr in the combined
2006-07 seasons. The difference in the field brightness between the ON and the
OFF field was also taken into account using a technique called {\sl library
padding} \cite{Scalzoetal2004}. After cuts and padding, a net
ON-source excess of 236 events was seen against a background of 5547
events. At a significance of $2.3\sigma$, this excess is insufficient to
claim a detection of 1ES 1218+304 but is used to establish flux upper
limits for the source. There were no significant transient events in
the data, as shown by the histogram of significances for each of the
152 pairs in Figure 2. 

\begin{figure}[b!]
\begin{center}
\includegraphics*[width=0.48\textwidth, angle=90,height=1.6in]{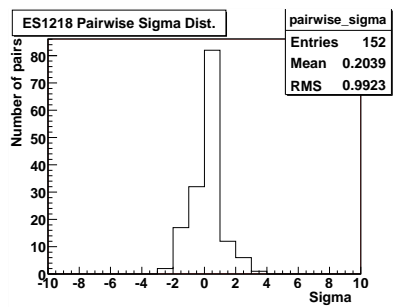}
\end{center}
\caption{Pair-wise sigma distribution for the 1ES 1218+304 data
  set.}\label{fig2} 
\end{figure}

\begin{table}

\begin{minipage}{0.48\textwidth}
\begin{center}
\caption{STACEE Data on 1ES 1218+304}
\vskip 0.05in
\begin{tabular}{llcc}\hline\hline

Year      &  Live-$^a$ & ON-Source     & Detection   \\
          &  time        & Excess Events & Sig.\\
\hline\hline

2006 & 12.0 & \multirow{2}{*}{236} & \multirow{2}{*} {2.3$\sigma$}\\

2007 & 16.3 &  &  \\
       	                 
\hline

\end{tabular}
\end{center}
Note. -- $a$ Total hours remaining after quality cuts. 
\end{minipage}

\end{table}

\section{Detector Simulations}

In order to understand the results of the observations of 1ES 1218+304, Monte Carlo
simulations were carried out to determine the effective area of the
STACEE detector, as described in \cite{Brameletal2005}. The effective
area was calculated as a function of energy at several hour angles,
for gamma-ray and proton primaries. Figure 3 shows the net effective 
area of STACEE for the 1ES 1218+304 observations. The simulated
effective area was then used to calculate the detector energy
threshold and fluxes. The energy threshold $E_{th}$ is
defined as the peak of the response curve obtained by multiplying the
detector effective area with the source spectrum. 
A spectral index of $-3.0$ was assumed for 1ES 1218+304, 
based on the measurement of the gamma-ray spectrum by MAGIC
\cite{Albert2006}. 
For 1ES 1218+304 we get $E_{th}$ $\sim 150\pm45_{\rm sys}$ GeV. 

\begin{figure}[b!]
\begin{center}
\includegraphics*[width=0.48\textwidth, angle=90,height=2in]{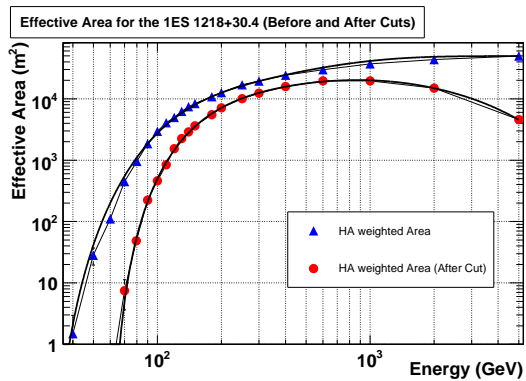}
\end{center}
\caption{Hour-angle weighted effective area of the STACEE detector for
  the 1ES 1218+304 observations. The lower plot shows the effective
  area after quality cuts. }\label{fig3}
\end{figure}

\section{Summary \& Discussions}

STACEE observed the high frequency-peaked BL Lac object 1ES 1218+304
in 2006 and 2007. After all cuts and padding 28.3 hr of data yielded
an ON-source excess with a significance of $2.3\sigma$ consistent with
no detected flux. Simulated effective areas (Figure 3) were used to
derive flux upper limits. For the combined 2006 and 2007 data sets the
differential flux upper limit at the 99\% confidence level was derived
to be $< 5.2 \times 10^{-6}$ m$^{-2}$ s$^{-1}$ TeV$^{-1}$ at 150
GeV, the energy threshold of STACEE. The upper limit was calculated assuming that the differential flux of photons follows a power law with an index of $-3.0$, as measured by MAGIC \cite{Albert2006}. The STACEE upper-limit is shown overlaid on the MAGIC spectrum in Figure 4. 
These numbers are consistent with the spectrum measured by MAGIC
\cite{Albert2006}, and with the recent VERITAS
results indicating a weak gamma-ray source at $\sim5$\% of the Crab
flux \cite{Fortin2007}. 

\begin{figure}
\begin{center}
\includegraphics*[width=0.48\textwidth,height=2.5in]{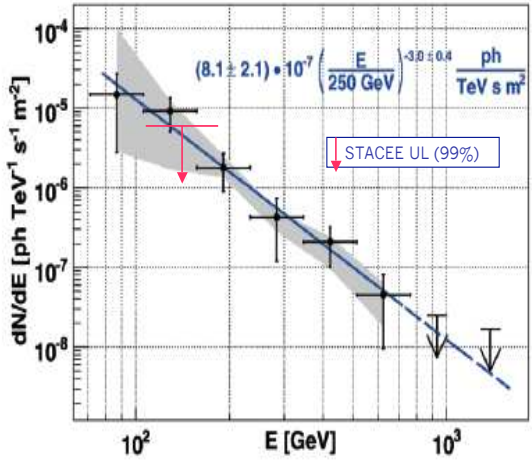}
\end{center}
\caption{Gamma-ray spectrum of 1ES 1218+304, as measured by MAGIC
  (figure from \cite{Albert2006}). The STACEE 99\% flux upper limit is overlaid on the plot at 150 GeV, the energy threshold of STACEE, as obtained from detector simulations. The upper limit was calculated assuming the power-law spectral index of $-3.0$ measured by MAGIC. } \label{fig4}
\end{figure}

\bigskip
\vskip 2in

{\bf Acknowledgements: }
Many thanks go to the staff of the National
Solar Tower Test Facility, who have made this work possible. This work
was funded in part by the US National Science Foundation, the Natural
Sciences and Engineering Research Council of Canada, Fonds Quebecois
de la Recherche sur la Nature et les Technologies, the Research
Corporation, and the University of California at Los
Angeles. R. M. acknowledges support from NSF grant 0601112.

\end{document}